\documentclass[10pt,conference]{IEEEtran}
\usepackage{epsfig}
\usepackage{amsmath}
\usepackage{amsfonts}
\usepackage{amssymb}
\usepackage[all]{xy}
\usepackage{xcolor}
\usepackage{graphicx}
 \DeclareGraphicsExtensions{.eps}

\newlength{\tmplength}
\newcommand{\tspace}[1][4.5mm]{\settoheight{\tmplength}{X}\addtolength{\tmplength}{#1}\hbox{\protect\raisebox{\tmplength}{}}}
\newcommand{\gf}{{\mathbb F}}%
\newcommand{\SR}{{\mbox{\rm\scriptsize SR}}}
\newcommand{\SD}{{\mbox{\rm\scriptsize SD}}}
\newcommand{\RD}{{\mbox{\rm\scriptsize RD}}}
\newcommand{\dx}{\mbox{\rm d}x}
\newcommand{\medoplus}{\mathop{\mbox{\LARGE $\oplus$}}}
\newenvironment{mylist}[1][$\bullet$]{\begin{list}{#1}{\leftmargin \parindent \itemindent 0mm \labelwidth \parindent}}{\end{list}}

\IEEEoverridecommandlockouts
\title{Split-Extended LDPC codes for coded cooperation}
\author{Valentin~Savin, CEA-LETI, MINATEC, Grenoble, France, valentin.savin@cea.fr%
\thanks{This work was partially carried out in the scope of Celtic CP5-026 (WINNER+) project.}}
\date{}
\begin{document}
\maketitle

\begin{abstract}
We propose a new code design that aims to distribute an LDPC code over a relay channel. It is based on a {\em
split-and-extend} approach, which allows the relay to split the set of bits connected to some parity-check of the LDPC
code into two or several subsets. Subsequently, the sums of bits within each subset are used in a repeat-accumulate
manner in order to generate extra bits sent from the relay toward the destination. We show that the proposed design
yields LDPC codes with enhanced correction capacity and can be advantageously applied to existing codes, which allows
for addressing cooperation issues for evolving standards. Finally, we derive density evolution equations for the
proposed design, and we show that Split-Extended LDPC codes can approach very closely the capacity of the Gaussian
relay channel.
\end{abstract}


\section{Introduction}
By exploiting the broadcast nature and the inherent spatial diversity of wireless communications, Sendonaris {\em et
al.} introduced the concept of {\em cooperative diversity} \cite{SendErkipAazh03:partI,SendErkipAazh03:partII} over
wireless relay channels and their multi-terminal extensions. A relay channel is a three terminal network consisting of
a source, a relay, and a destination. The source broadcasts a message to both relay and destination, while the relay
forwards the message to the destination. Subsequently, many authors proposed cooperation protocols for the relay
channel, which can be classified into two major categories, namely the amplify-and-forward (AF) and the
decode-and-forward (DF) \cite{LaneTseWorn04}. In AF protocols, the relay simply amplifies the received signal and
forwards it to the destination. The DF protocol allows the relay to decode the received signal, re-encode it, and
forward it to the destination. The forwarded message can either be identical to, or part of the initial transmission
(repetition coding), or it can be obtained by using a dedicated coding scheme at the relay (distributed coding). In the
first case the destination combines received signals from both source and relay, which results in an improved
signal-to-noise ratio (SNR) on the received transmission. Besides, the same code is used for encoding at the source and
decoding at the destination. In the second case, the destination gains knowledge of extra information, but it needs a
dedicated decoding scheme, able to jointly decode received signals from both source and relay.

One of the most known examples of distributed coding is the one of a distributed turbo-code \cite{ValenZhao03:DTC}: the
source broadcasts a recursive convolutional code (RCC) to both relay and destination. After decoding, the relay
interleaves and re-encodes the message using the same RCC, prior to forwarding it to the destination. Because the
destination receives both codes in parallel, it can jointly decode received signals from source and relay by using a
parallel-concatenated turbo-code.

Low-Density Parity-Check (LDPC) codes play a prominent role in the family of error-correcting codes. They feature low
complexity decoding and can be optimized for a broad class of channels, with performance approaching the theoretical
Shannon limit \cite{Rich-Shok-Urba}. Although they lend themselves less easily to distributed schemes, several
approaches have been already proposed in the literature
\cite{KhoAhmAaz04,RazYu06:bilayerLDPC,RazYu07:bilayerLDPC,ChakBaynSabhAazh07,HuDuman07}, %
\cite{LiYueEtAl08:ldpc_relay_syst,CancMegh09,DuyckBoutMoen09}. 
Some of these approaches are somehow based either on serial or parallel code concatenation\footnote{Meaning that the
graph of the LDPC code broadcasted from the source is a subgraph of the destination decoding graph.}, or on punctured
(rate-compatible) LDPC codes. From the code design point of view, the serial or parallel concatenation of LDPC codes
has intrinsic limitations, mainly because parity-check matrices used for decoding at the relay and at the destination
are included one in the other, resulting in inappropriate matrix topologies (density on non-zero entries, column and
row weight distributions, cycles, etc.). Punctured codes also present some weaknesses in the context of coded
cooperation. In such a cooperation scheme, punctured bits, which are not broadcasted by the source, are transmitted
from the relay toward the destination. Hence, the punctured code has to be robust, such as to allow successful decoding
at the relay: in practice punctured bits are those bits which are {\em the easiest to retrieve} by the iterative
decoding process. On the other hand, bits transmitted from the relay to the destination encounter better channel
conditions: advantageously, these bits should be those which are {\em the most difficult to retrieve} by the iterative
decoding process. The contradiction between these two requirements leads to an imbalance in the design of the
puncturing pattern.

In this paper we propose a new code design method that aims to create incremental redundancy for LDPC codes, while
avoiding both code concatenation and code puncturing. It is based on a {\em split-and-extend} approach, which can be
seen as the ``coding analogous of the {\em divide-and-conquer} concept''. After decoding the received signal, the relay
computes extra parity bits by splitting parity-checks of the initial code. Hence, each extra parity bit is the sum of
some subset of bits participating in the same parity-check of the initial code. Then the relay transmits these new
parity bits towards the destination. The whole process amounts to create a new matrix, whose rows correspond to
parity-checks involving both old and new parity bits. These parity-checks are therefore distributed over the relay
channel, in the sense that part of checked bits are received on the source-to-destination link, and another part are
received on the relay-to-destination link. Consequently, this new matrix can be used at the destination to jointly
decode the received signals from both the source and relay.

The paper is organized as follows. The proposed code design is introduced in Section \ref{sec:seldpc}. In Section
\ref{sec:backward_comp} we show that the proposed design can be advantageously applied to to existing codes, which
allows for backward compatibility while addressing cooperation issues for evolving standards.
In Section \ref{sec:asymptotic_opt} we propose a ``coding-perspective'' analysis of cooperative systems. We introduce
threshold and capacity functions, and we derive density evolution equations for Split-Extended (SE) LDPC codes over the
Gaussian relay channel.

Numerical results are shown in Section \ref{sec:num_results}, and Section \ref{sec:conclusions} concludes this paper.

\section{Split-Extended LDPC codes} \label{sec:seldpc}

\subsection{Basic idea}
\begin{figure}[!b]
\includegraphics*[height=\linewidth,angle=-90]{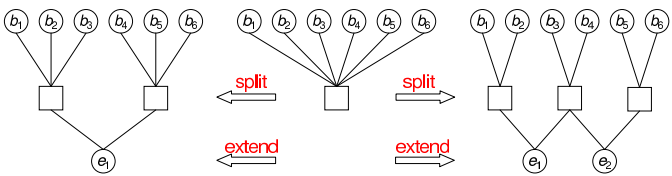}
 \caption{Split-Extension examples}\label{fig:split_extend}
\end{figure}

The basic idea of the split-extend design can be resumed as follows. Let $H_1$ be the parity-check matrix of the LDPC
code broadcasted by the source to both relay and destination. Hence, broadcasted bits satisfy parity-check equations
corresponding to the rows of $H_1$. After decoding the received signal, the relay computes extra parity bits by
splitting these parity-checks, as illustrated at Figure~\ref{fig:split_extend}. The parity-check in the middle
corresponds to a row of $H_1$. In the left example, a new parity bit $e_1$ is created by spitting the original
parity-check into two sub-checks. Precisely, this means that the set of bits connected to the check-node is partitioned
into two subsets, and the parity bit $e_1$ is generated as the sum of the bits of either one of the two subsets. In the
right example, two new parity bits $e_1$ and $e_2$ are created by spitting the original parity-check into three
parity-checks. Precisely, the set of bits connected to the check-node is partitioned into three subsets, and $e_1$ is
generated as the sum of the bits in the first subset. Subsequently, $e_2$ can be generated either as the sum of $e_1$
and the bits in the second subset, or as the sum of the bits in the third subset. The total number of extra parity bits
depends on the number of rows of $H_1$ and the number of extra bits generated for each row of $H_1$ (which may vary
from one row to another). The sequence of all the extra parity bits, denoted by $E = (e_1, e_2, \dots)$, is then
transmitted from the relay to the destination. The matrix $H$ obtained by the split-extension of $H_1$ (i.e. the
incidence matrix of the split-extended graph) verifies $H\cdot(X, E)^t = 0$, where $X$ denotes the codeword broadcasted
by the source. Therefore, $H$ can be used at the destination in order to jointly decode the received signals from both
source and relay.

\begin{figure}[!b]
\includegraphics*[height=\linewidth,angle=-90]{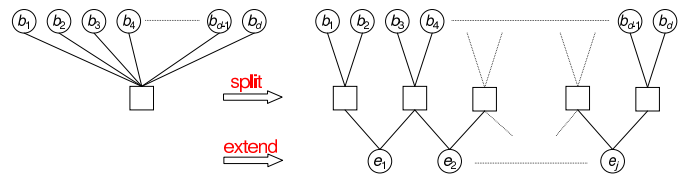}
 \caption{Repeat-Accumulate Split-Extension}\label{fig:split_extend_theta}
\end{figure}

A more general example of split-extension is illustrated at Figure~\ref{fig:split_extend_theta}. The original
check-node is split into several sub-checks, and extended bits are generated in a repeat-accumulate manner. Such a
split-extension will be referred to latter in the paper (Section~\ref{sec:asymptotic_opt}) as {\em repeat-accumulate
split-extension}.

\subsection{The general case}
\noindent The following notation will be used throughout this section:
\begin{mylist}
\item For any positive integer $N$, $[1:N]=\{1,\dots,N\}$ denotes the set of integers between $1$ and $N$, inclusive.
\item For any subset $S\subset[1:N]$, $[1:N]\setminus S$ denotes the set of integers
between $1$ and $N$ that are not in $S$.
\item For any length-$N$ vector $V = (v_1, \dots, v_N)$ and any subset $S=\{i_1,\dots,i_k\}\subset[1:N]$, $V|_S$
denotes the length-$k$  vector defined by the coordinates of $V$ which are in $S$, that is $V|_S = (v_{i_1},\dots,
v_{i_k})$
\item For any binary matrix $H$ of size $M\times N$, ${\cal R}(H) \subset \{0,1\}^N$ denotes the set of the $M$
row-vectors of $H$. A partition of ${\cal R}(H)$ is a set of nonempty subsets ${\cal P}_1,\dots,{\cal P}_{M_1}
\subseteq {\cal R}(H)$ such that every row of $H$ is in exactly one of these subsets.
\item For any binary vectors $V_1, V_2\in\{0,1\}^N$, $V_1\oplus V_2$ denotes their componentwise sum modulo $2$.
\end{mylist}

\smallskip\noindent{\em [Definition]} Let $H_1$ and $H$ be two parity-check matrices of size $M_1\times N_1$
and  $M\times N$, respectively, with $N_1\leq N$. We say that the matrix $H$ is obtained by {\em split-extending} the
matrix  $H_1$ if there exist a partition of the $M$ rows of  $H$ in $M_1$ disjoint subsets ${\cal P}_1,\dots,{\cal
P}_{M_1} \subseteq {\cal R}(H)$, and a subset $S\subset[1:N]$ with cardinality $N_1$, such that for any $m\in[1:M_1]$:
  $$ \medoplus_{L\in{\cal P}_m} L|_S = R_m \mbox{ and }  \medoplus_{L\in{\cal P}_m} L|_{[1:N]\setminus S} = 0,$$
where $R_1,\dots,R_{M_1}$  denote the $M_1$  rows of $H_1$. In this case, the set $E = [1:N]\setminus S$ is called {\em
set of extended bit-nodes}. The split-extension is called {\em non-singular} if the columns of $H$ corresponding to
 $E$ are linearly independent.

\smallskip
Now, consider two parity-check matrices $H_1$ and $H$, such that $H$ is a non-singular split-extension of $H_1$. Then:
\begin{mylist}
\item If $X$ is a codeword\footnote{By abusing language,
we say that $X$ is a codeword of $H$, if $HX^t = 0$}  of $H$, then $X|_S$ is a codeword of $H_1$
\item For any codeword $X_1$ of $H_1$, there exists a unique codeword $X$ of $H$, such that $X|_S = X_1$
\end{mylist}

Matrices ($H_1$, $H$) can be used within a cooperative transmission system as follows:
\begin{mylist}
\item The source encodes the
packet of information bits, generating a codeword $X_1$  of  $H_1$. It broadcasts $X_1$  to both relay and destination.
\item The relay decodes the received signal, correcting the transmission errors on  $X_1$.
It generates a codeword $X$ of  $H$, such that  $X|_S = X_1$, and sends the set of extended bits $X|_{E}$ towards the
destination.
\item Thus, the destination receives noisy versions of $X|_S = X_1$ and $X|_{E}$
(from both the source and the relay), which can be decoded using the matrix $H$.
\end{mylist}

\section{Split-extend design for backward compatibility}\label{sec:backward_comp}
\begin{figure}[!h]
\noindent\includegraphics*[height=.7\linewidth,angle=-90]{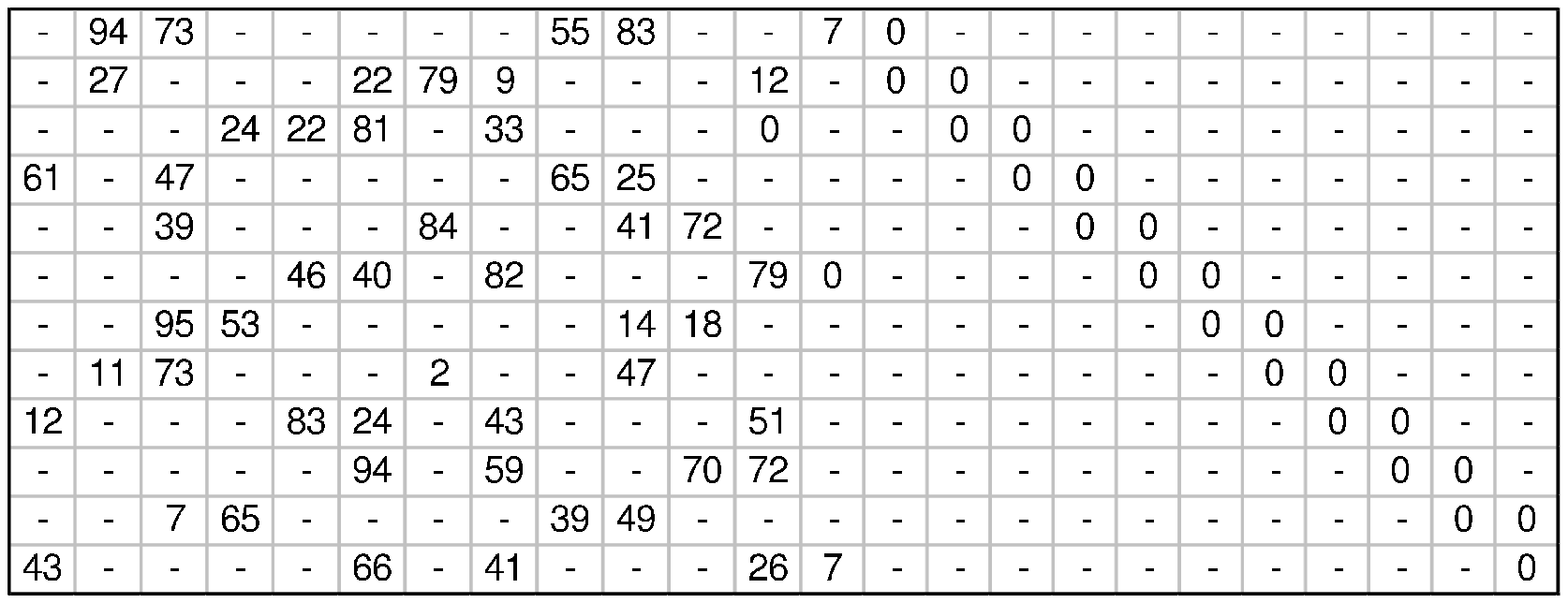}

\smallskip
\noindent\includegraphics*[height=\linewidth,angle=-90]{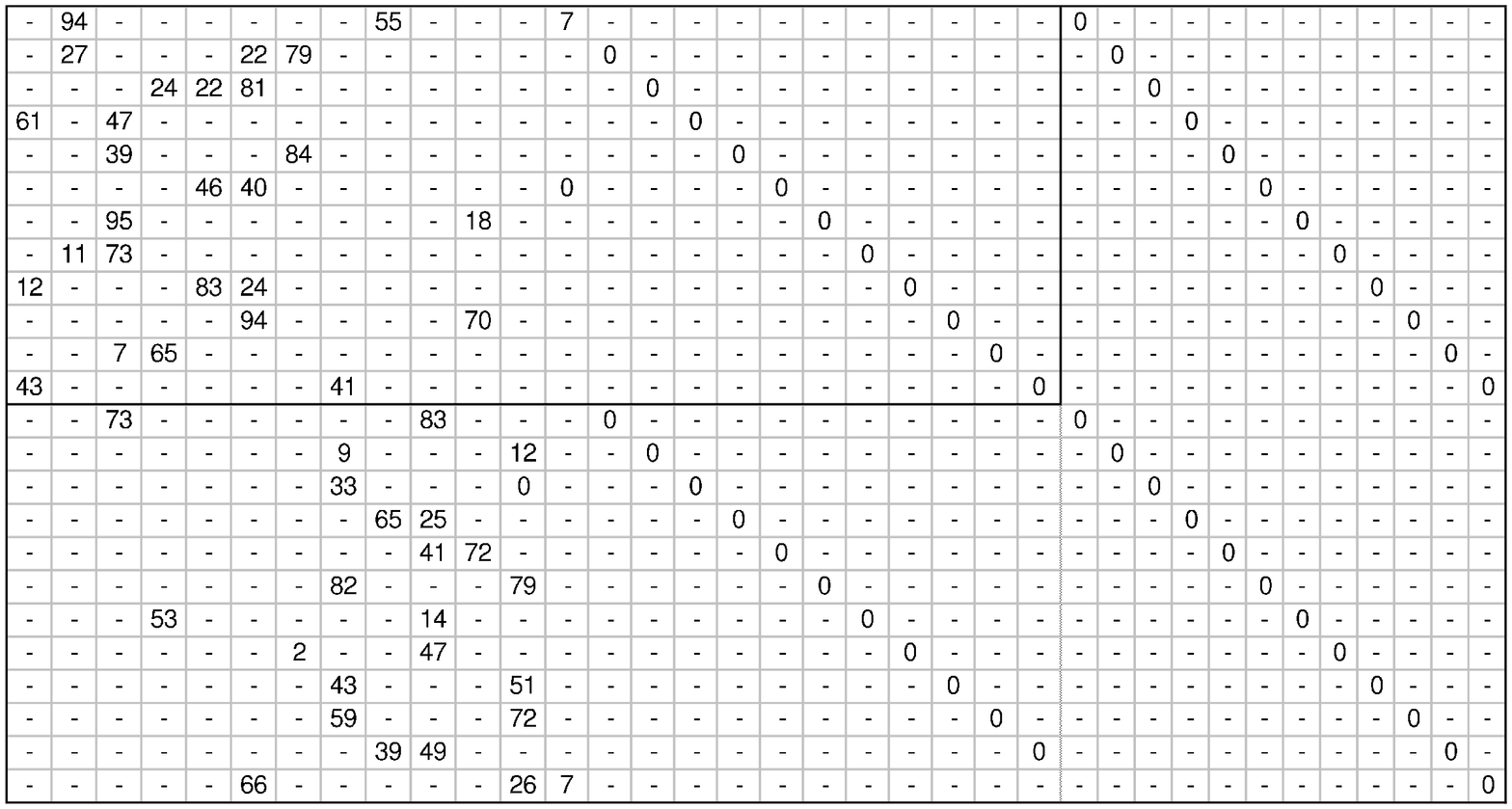}

\caption{Base matrix of the QC-LDPC WiMAX code with rate 1/2 (top), and corresponding split-extended base matrix
(bottom)} \label{wimax_se_rate12} 
\end{figure}

This section is independent of the following sections, though, it highlights an interesting property of the proposed
design: it can be advantageously applied to existing codes, which allows for addressing cooperation issues for evolving
standards, while maintaining backward compatibility with a reduced impact on user equipment. To illustrate this, the
LDPC codes from the IEEE.802.16e (WiMAX) standard \cite{wimax_std} with coding rates $1/2$ and $2/3$ have been
split-extended, such that the number of generated extended bits be equal to the number of information bits. Thus, for
coding rate $1/2$, each row of the parity-check matrix has been split into two rows; while for rate $2/3$, each row of
the parity-check matrix has been split into three rows (see Figure~\ref{fig:split_extend}). Splitting has been
performed by a dedicated algorithm that search for short cycles in the parity check matrix, then splits rows such that
to break as many short cycles as possible. Base matrix of the Quasi-Cyclic (QC) LDPC WiMAX code with rate 1/2 and the
corresponding split-extended matrix are shown at Figure \ref{wimax_se_rate12} ($-1$'s entries of the base matrix are
represented by a dash sign).

Clearly, split-extended matrices can be used to address cooperation issues for uplink transmissions, in a completely
transparent way for the user: the user encodes the transmitted signal by using the original parity check matrix; the
relay decodes the signal, then computes and sends the sequence of extended parity bits to the base station, which will
use the split-extended matrix in order to decode the received signals from both user and relay. Hence, split-extended
matrix is only needed at the relay and the base station. For downlink transmissions, if the user terminal is equipped
with split-extended matrices, the situation is symmetric. Otherwise, the relay can only repeat the sequence of
information bits, which, however, provides the user with an energy gain on the information sequence.

\begin{figure}[!h]
\vspace{-5mm} \noindent\includegraphics[width=\linewidth]{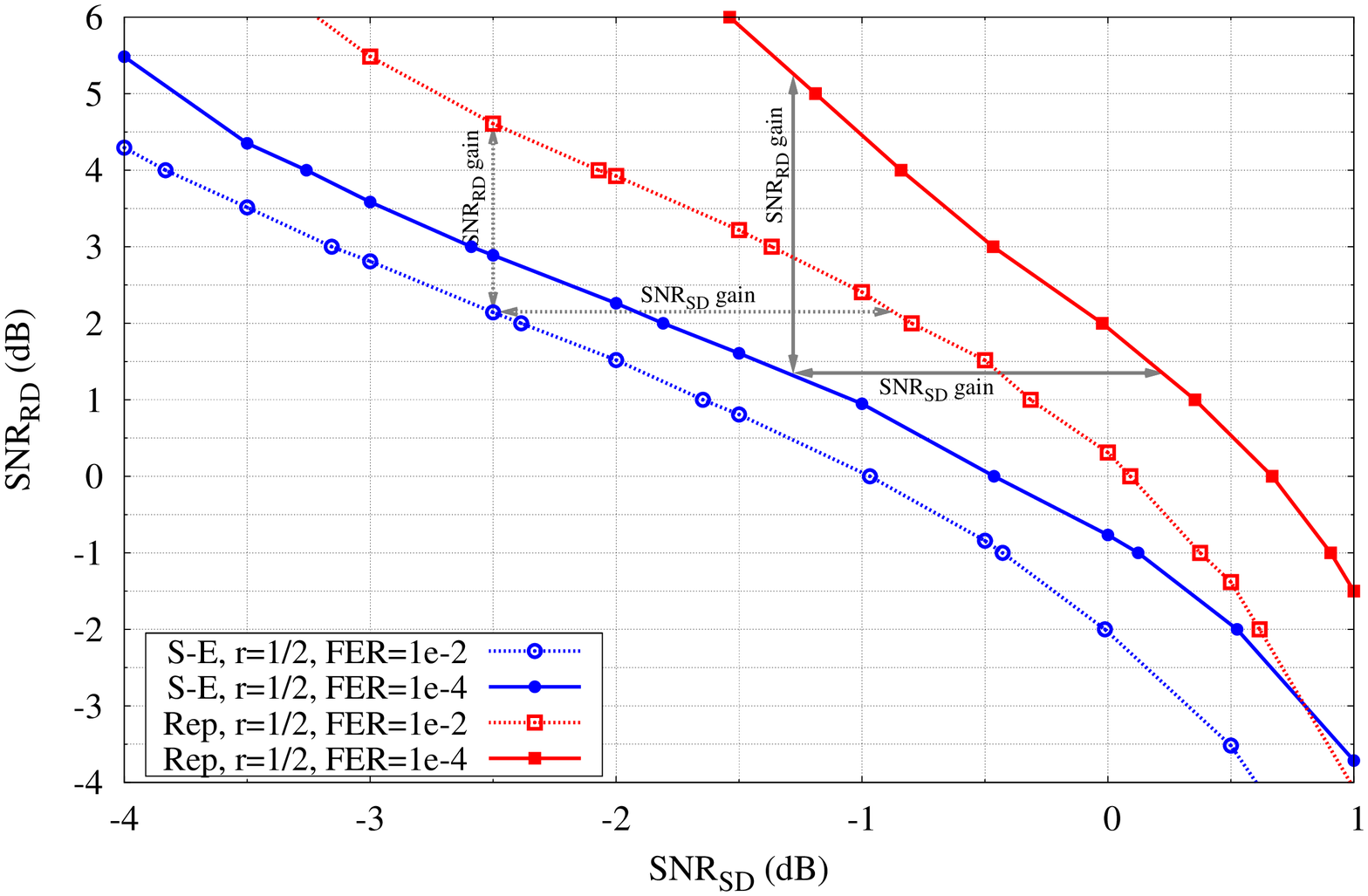}

\vspace{-5mm}
 \caption{Split-Extend vs. Repetition coding for WiMAX code with rate $1/2$} \label{tgtFER_r12} 
\end{figure}

\begin{figure}[!h]
\vspace{-5mm} \noindent\includegraphics[width=\linewidth]{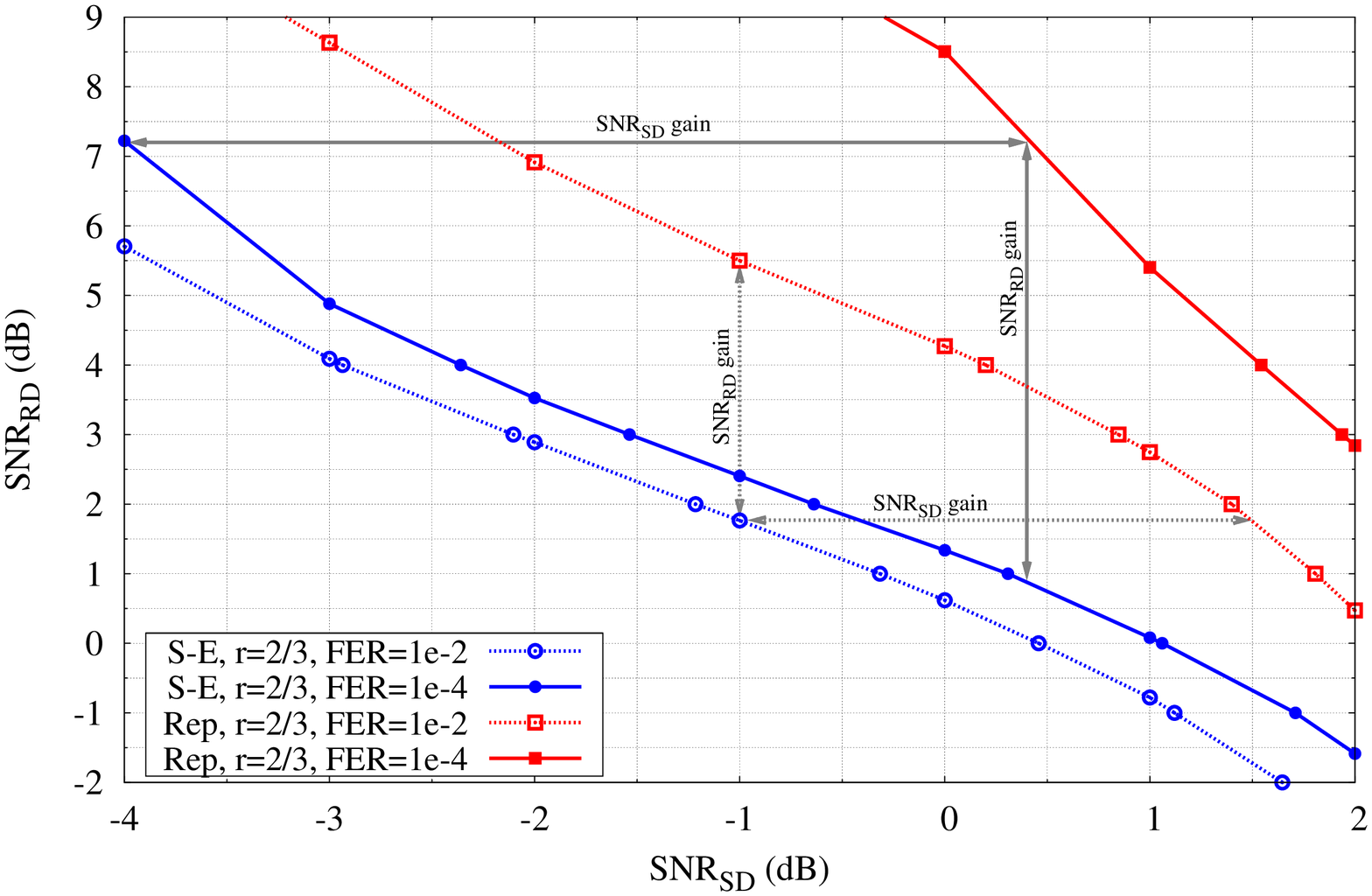}

\vspace{-5mm}
 \caption{Split-Extend vs. Repetition coding for WiMAX code with rate $2/3$} \label{tgtFER_r23} \vspace{-5mm}
\end{figure}

Simulation results over the AWGN relay channel, with QPSK modulation, are shown at Figures \ref{tgtFER_r12} and
\ref{tgtFER_r23}. The source broadcasts either the WiMAX code with rate $1/2$ (Fig. \ref{tgtFER_r12}) or the WiMAX code
with rate $2/3$ (Fig. \ref{tgtFER_r23}), and the SNR on the source-to-relay link is fixed to $2.5$ and $4.5$ dB,
respectively. Both figures compare the performance of two cooperation scenarios: the relay generates and transmits
extended bits in the first scenario, while in the second, it forwards the (error-corrected) sequence of information
bits. Plotted curves represent SNRs required on source-to-destination and relay-to-destination links, such that to
obtain a target frame error rate either of $10^{-2}$ (dotted curves) or of $10^{-4}$ (solid curves). The Self-Corrected
Min-Sum algorithm \cite{Savin08:SCMS} is used for decoding at both relay and destination.
The gap between the dotted and solid curves is determined by the slope of the frame error rate curves in the waterfall
region. However, for the second scenario (repetition of the information sequence) the increased gap between the two
curves for small $\mbox{SNR}_{\SD}$ values is also justified by a frame error rate error floor  above  $10^{-4}$. Note
also \hfill that \hfill only \hfill SNR \hfill pairs \hfill with \hfill $\mbox{SNR}_{\RD} > \mbox{SNR}_{\SD}$ \hfill
are \hfill likely \hfill to

\noindent  be encountered in practice. %
The SNR gain between the two scenarios can be measured either as the horizontal distance (for the source-to-destination
link) or the vertical distance (for the relay-to-destination link) between corresponding curves. It can be observed
that split-extended codes achieve a significant SNR gain, in order of several dBs, over the repetition scenario.

\section{Asymptotic analysis of SE-LDPC codes}\label{sec:asymptotic_opt}
We denote by ${\cal E}(\lambda, \rho)$ the ensemble of LDPC codes with edge-perspective degree distribution polynomials
$\lambda$ and $\rho$ \cite{Rich-Urba}. It is well known that when the code length tends to infinity, (almost) all the
codes of the family behave alike, and they exhibit a threshold phenomenon, separating the region where reliable
transmission is possible from that where it is not \cite{Rich-Urba}.

Consider some channel model depending on a parameter $\sigma$, such that the channel conditions worsen when $\sigma$
increases (for instance, the noise variance for the AWGN channel, or the error probability for the BSC channel). The
{\em threshold}\break of the ensemble ${\cal E}(\lambda, \rho)$ is defined as the supremum value of $\sigma$ (worst
channel condition) that allows transmission with an arbitrary small error probability, assuming that the transmitted
data is encoded with an arbitrary-length code of ${\cal E}(\lambda, \rho)$.

The above threshold phenomenon can be extended to the relay channel, but the following 
must be taken into account: 
\begin{mylist}
\item the channel is modeled by three parameters $\sigma_\SR$, $\sigma_\SD$, and $\sigma_\RD$,
corresponding, with obvious notation, to the three links between source, relay, and destination.
\item the ensemble of SE-LDPC codes depends not only on $\lambda$ and $\rho$, but also on the {\em splitting distribution}.
\end{mylist}

\noindent{\em [Channel assumptions]} Since we are strictly interested on code analysis, the following assumptions will
be made:
\begin{mylist}
  \item when the relay fails to decode the received signal from the source, it does not transmit any information
  to the destination,
  \item the relay channel is degraded, in the sense that the above parameters must satisfy
  $ \sigma_\SR < \sigma_\SD$ and $\sigma_\RD < \sigma_\SD$.
\end{mylist}

\smallskip
\noindent{\em [Distributed code]} A linear distributed code of dimension $K$ is a vector subspace ${\cal C}\subset
\gf_2^{N_1}\times\gf_2^{N_2}$, such that ${\cal C}$ and its projection on $\gf_2^{N_1}$ are both of dimension $K$. The
{\em distributed rate} of ${\cal C}$ is by definition $(r_1, r_2) = (\frac{K}{N_1}, \frac{K}{N_2})$. Hence, $r_1\leq
1$, but $r_2$ can be greater than $1$. The overall coding rate is defined as $r = \frac{K}{N_1+N_2}=\frac{r_1
r_2}{r_1+r_2}$. The idea behind is that the first $N_1$ bits of a codeword $c\in{\cal C}$ are broadcasted from the
source to both relay and destination and, in case that the relay manages to decode the received signal\footnote{Thus,
this definition is dependent on the above channel assumptions.}, it transmits the last $N_2$ bits toward the
destination.

\smallskip
\noindent{\em [SE-LDPC ensembles]} Let $H_1$ be the parity-check matrix of the LDPC code broadcasted by the source, and
let $\theta \geq 2$. For each parity-check of $H_1$, assume that:
\begin{mylist}
 \item the set of bits connected to the parity-check is partitioned into $\theta$ subsets of (almost) regular size;
that is, each subset contains either $\left\lfloor\frac{d}{\theta}\right\rfloor$ or
$\left\lceil\frac{d}{\theta}\right\rceil$ bits participating in the parity-check, where $d$ denotes the parity-check
degree,
 \item extended bits are generated in a repeat-accumulate manner (Figure~\ref{fig:split_extend_theta}).
\end{mylist}
 The resulting distributed SE-LDPC code will be referred to as having {\em splitting degree} $\theta$. We
denote by ${\cal E}_\theta(\lambda, \rho)$ the ensemble
of repeat-accumulate SE-LDPC with splitting
degree $\theta$, obtained by split-extending LDPC codes with edge-perspective degree distribution polynomials $\lambda$
and $\rho$.

\subsection{SE-LDPC ensemble thresholds}
Given an ensemble of SE-LDPC codes, our intention is to separate the region of parameters $(\sigma_\SR, \sigma_\SD,
\sigma_\RD)$ where reliable transmission is possible from that where it is not.

Let $\sigma^{*}_1(\lambda,\rho)$ be the threshold of the ${\cal E}(\lambda, \rho)$ ensemble. If $\sigma_{\SR} >
\sigma^{*}_1(\lambda,\rho)$, the error decoding probability at the relay is lower bounded by a positive constant.
Consequently, reliable cooperation cannot be achieved, as the relay does not transmit any information to the
destination when it fails to decode the received signal, and the destination cannot reliably decode the signal received
from the source, since $\sigma_\SD > \sigma_\SR > \sigma^{*}_1(\lambda,\rho)$.

From now on we consider that $\sigma_{\SR} < \sigma^{*}_1(\lambda,\rho)$; hence, we may assume {\em error
free}\footnote{With arbitrarily small error probability, as the code length tends to infinity.}
transmission between source and  relay. We will also use the
following notation:
\begin{mylist}
\item $\sigma = \sigma_{\RD}$ (we drop subscript $\RD$), which will be referred to as {\em noise parameter},
\item $\delta = \displaystyle\frac{\sigma_\SD}{\sigma_\RD}\geq
  1$, which will be referred to as {\em channel discrepancy}.
\end{mylist}

The {\em threshold function} $\sigma^{*}_{\theta,\lambda,\rho}:[1,\infty[\rightarrow \mathbb{R}$ associates with each
discrepancy value $\delta$ the  noise  threshold $\sigma^{*}_{\theta,\lambda,\rho}(\delta)$, defined as the supremum
value of $\sigma$ that allows transmission with an arbitrary small error probability, assuming that the transmitted
data is encoded with an arbitrary-length distributed code from ${\cal E}_\theta(\lambda, \rho)$. This definition makes
sense only under the implicit assumption of a concentration theorem as in \cite{Rich-Urba}, which can indeed be derived
by using the same arguments as in {\em loc. cit.}

The threshold function can be efficiently computed by tracking the density evolution of messages exchanged within the
iterative decoding, as explained in the next section.

\subsection{Density evolution}
Throughout this section, we assume binary-input AWGN relay channel. We combine a multi-edge approach
\cite{Rich-Urba:multiedge} and the Gaussian approximation
method proposed in \cite{Chung-Richar-Urba:DE_GA}, in
order to derive density evolution equations for the SE-LDPC code ensemble ${\cal E}_{\theta}(\lambda,\rho)$. We
separate the set of bit-nodes of the expanded graph into two subsets:
\begin{mylist}
\item type-$1$ bit-nodes, which correspond to bits received by the destination from the source,
\item type-$2$ bit-nodes, which correspond to extended bits received by the destination from the relay.
\end{mylist}
We distinguish between type-$1$ and type-$2$ edges, according to whether they are incident to type-$1$ or type-$2$
bit-nodes. Moreover, check-node degrees are also defined type-wise. Hence, we say that a check-node is of degree $(d_1,
d_2)$ if it is connected to $d_1$ type-$1$ bit-nodes and $d_2$ type-$2$ bit-nodes. From our definition of SE-LDPC
ensembles, it follows that the type-$2$ degree $d_2$ is equal either to $1$ or $2$ (see Fig.
\ref{fig:split_extend_theta}). Finally, for each type $t=1,2$, we define:

\begin{mylist}
  \item $\lambda^{[t]}_d$ is the fraction of
  type-$t$ edges connected to bit-nodes of degree $d$,
  \item $\rho^{[t]}_{d_1, d_2}$
  is the fraction of type-$t$ edges connected to check-nodes of degree $(d_1, d_2)$.
\end{mylist}

\noindent It follows that $\lambda^{[1]}_d = \lambda_d$, $\lambda^{[2]}_2 = 1$ ($\lambda^{[2]}_d = 0$ for
$\mbox{\small$d\neq 2$}$), while $\rho^{[1]}_{d,i}$ and
$\rho^{[2]}_{d,i}$ ($\mbox{\small$i=1,2$}$) can be computed as follows:  {
$$\begin{array}{r@{\ }c@{\ }l@{\ \ \ \ \ }r@{\ }c@{\ }l}
  \rho^{[1]}_{d,1} & = & \displaystyle d \sum_{j=-\theta+1}^{\theta-1}
  k_{j,1} \frac{\rho_{d\theta+j}}{d\theta+j}, &  \rho^{[2]}_{d,1} & = &
  \displaystyle  \frac{\bar\rho \rho^{[1]}_{d,1}}{2d(\theta-1)}, \\
  \rho^{[1]}_{d,2} & = & \displaystyle d \sum_{j=-\theta+1}^{\theta-1} k_{j,2}
   \frac{\rho_{d\theta+j}}{d\theta+j}, &  \rho^{[2]}_{d,2} & = &
  \displaystyle  \frac{\bar\rho \rho^{[1]}_{d,2}}{d(\theta-1)},
\end{array}$$
}

\noindent where $\mbox{\small$\bar\rho = \displaystyle \frac{1}{\int_0^1 \rho(x) \,\dx}$}$ is the average check-node
degree of the original (unsplit) graph, and {\small
 $$\begin{array}{r@{\ }c@{\ }l}
   k_{j,1} & = & \left\{\begin{array}{ll}
             0, & \mbox{ if } \ 2 \leq j \leq \theta-1 \\
             1, & \mbox{ if } \ j = 1  \mbox{ or } j=-\theta+1 \\
             2, & \mbox{ if } \ -\theta+1 < j \leq 0
   \end{array}\right.\\
   k_{j,2} \tspace[2mm]& = & \theta-|j|-  k_{j,1}
   \end{array}$$}

Now, under the Belief-Propagation decoding, let $m_{v^{[t]}}^{(\ell)}$ denote the mean of outgoing messages from
type-$t$ bit-nodes at iteration $\ell$. Let also $r^{[t]}_\ell = 1 -
E\left(\tanh\frac{m_{v^{[t]}}^{(\ell)}}{2}\right)$, where $E$ denotes as usual the expected value operator. Define:

{\footnotesize
$$\begin{array}{r@{\ }c@{\ }l}
  \displaystyle \phi(x) & = & \displaystyle 1 - \frac{1}{2\sqrt{(\pi x)}}\int_{\mathbb{R}}
  \tanh\frac{u}{2}e^{-\frac{(u-x)^2}{4x}} \mbox{d}u, \ (\phi(0)=1) \\
  \psi(x) & = & \phi^{-1}(1-x) \\
  \displaystyle h^{[1]}(x,y) & = &
  \displaystyle \sum_j \lambda^{[1]}_j \phi\left(\frac{2}{(\delta\sigma)^2} + (j-1)
           \sum_{i_1,i_2} \rho^{[1]}_{i_1,i_2}\psi(x^{i_1-1}y^{i_2}) \right) \\
  \displaystyle h^{[2]}(x,y) & = &
  \displaystyle \sum_j \lambda^{[2]}_j \phi\left(\frac{2}{\sigma^2} + (j-1)
           \sum_{i_1,i_2} \rho^{[2]}_{i_1,i_2}\psi(x^{i_1}y^{i_2-1}) \right)
\end{array}$$
}

Then, under the assumption that the  messages exchanged during the iterative Belief-Propagation decoding are
independent and symmetric Gaussian distributed, $r^{[1]}_\ell$ and $\ r^{[2]}_\ell$ can be recursively computed by:
$$\mbox{\footnotesize$\displaystyle \left(r^{[1]}_\ell, \ r^{[2]}_\ell\right) = \left(h^{[1]}(1-r^{[1]}_{\ell-1},1-r^{[2]}_{\ell-1}), \ \ h^{[2]}(1-r^{[1]}_{\ell-1},1-r^{[2]}_{\ell-1})\right)  $},$$
with initial values {\small$\left(r^{[1]}_0, \ r^{[2]}_0\right) = \left(\phi\left(\frac{2}{(\delta\sigma)^2}\right),\
\phi\left(\frac{2}{\sigma^2}\right)\right)$}. The proof will be omitted, since it follows from the same arguments as in
\cite{Chung-Richar-Urba:DE_GA}. The above recursion holds as long as $\ell$ is less than half the girth of the graph,
which goes to infinity with the code-length, and the successful decoding condition for an ``infinite'' code from ${\cal
E}_\theta(\lambda, \rho)$ can be expressed as {\small$\left(r^{[1]}_\ell, \ r^{[2]}_\ell\right) \rightarrow 0$}.
Therefore, the threshold function defined in the above section, can be computed by:
$$\sigma^{*}_{\theta,\lambda,\rho}(\delta) = \sup\{\sigma \mid \lim_{\ell\rightarrow\infty} r^{[1]}_\ell
 = \lim_{\ell\rightarrow\infty} r^{[2]}_\ell = 0\}$$

\subsection{Theoretical limit}
In order to evaluate the performance
of an ensemble of codes, we would like to compare its threshold function with the {\em capacity function}, inferred
from the channel capacity.

Capacities of various relaying strategies have been computed in \cite{CoveGama79,KhojSabhAazh03,KramGastGupt05}, and
depend on the capacities of the three links. Since we assumed that source-to-relay transmission is error
free\footnote{In practice, the channel need not be error free; the assumption is that the $\sigma_\SR$ noise  is below
the threshold of the code broadcasted by the source.}, we only consider the two other links. Let $\gamma_\RD(\sigma)$
denote the information rate capacity of the relay-to-destination channel with parameter $\sigma$, and let
$\gamma_\SD(\sigma)$ be defined in a similar manner. The information rates are considered by transmitted bit, thus both
$\gamma_\RD(\sigma), \gamma_\SD(\sigma)\in[0,1]$. We also assume that the noise parameter $\sigma\in[0, +\infty[$ and
$\gamma_\RD$, $\gamma_\SD$ are continuous decreasing functions, such that $\gamma_\RD(0) = \gamma_\SD(0) = 1$ and
$\displaystyle\lim_{\sigma\rightarrow +\infty}\gamma_\RD(\sigma) = \lim_{\sigma\rightarrow +\infty}\gamma_\SD(\sigma) =
0$.

Now, let $(r_1, r_2)\in[0,1]\times[0,+\infty[$ be a target distributed coding rate. The capacity function
$\gamma_{r_1,r_2}:[1,+\infty[ \rightarrow [0,+\infty[$ is defined by  $\gamma_{r_1,r_2}(\delta) = \sigma$, where
$\sigma$ is the unique solution of the equation:
$$\frac{\gamma_\SD(\delta\sigma)}{r_1}+\frac{\gamma_\RD(\sigma)}{r_2} = 1$$
Note that for $\sigma=0$, we have $\frac{\gamma_\SD(0)}{r_1}+\frac{\gamma_\RD(0)}{r_2} = \frac{1}{r_1}+\frac{1}{r_2}
\geq \frac{1}{r_1} \geq 1$, and $\displaystyle\lim_{\sigma\rightarrow
+\infty}\mbox{$\frac{\gamma_\SD(\delta\sigma)}{r_1}+\frac{\gamma_\RD(\sigma)}{r_2}$} = 0$, thus such a solution always
exists and it is unique, due to the above assumptions.

The meaning of the capacity function is the following. Assume that we want to transmit information with distributed
rate $(r_1, r_2)$ over some relay channel. The rate $r_1$ is chosen according to the quality of the channel between
source and relay, such that to ensure error free transmission between them. The rate $r_2$ is generally chosen
according to the delay constraints of the cooperation system. The question is to know is there exists a distributed
code with distributed rate $(r_1,r_2)$ allowing error free\footnote{Arbitrary small error probability when the code
length goes to infinity.}  transmission. The answer is as follows. If the discrepancy and noise parameters $(\delta,
\sigma)$ of the relay channel verify $\sigma < \gamma_{r_1,r_2}(\delta)$ then such a distributed code exits. However,
note that a code allowing error free transmission for some pair $(\delta, \sigma)$, might not be suitable for some
other pair of parameters satisfying the above condition. If $\sigma
> \gamma_{r_1,r_2}(\delta)$, then reliable transmission with distributed rate $(r_1, r_2)$ is not possible. The proof
will be given in an extended version of this paper.

\section{Numerical results}\label{sec:num_results}
We assume BI-AWGN relay channel throughout this section. The following degree distribution pair, with designed coding
rate $1/2$, was designed by exact density evolution, and its threshold\footnote{Note that the threshold calculated by
Gaussian approximation is $0.9459$.} over the BI-AWGN channel is $\sigma^{*} = 0.9649$ \cite{Rich-Shok-Urba}.
{\small $$\begin{array}{@{\ }r@{\,}c@{\,\,}l}
 \lambda(x) & = & 0.2199x + 0.2333x^2 + 0.0206x^3 +0.0854x^5 + 0.0654x^6 \\
            &   &  + 0.0477x^7 + 0.0191x^8 + 0.0806x^{18}+0.2280x^{19} \\
 \rho(x)    & = & 0.6485x^7 + 0.3475 x^8 + 0.0040 x^9
\end{array}$$ }

\begin{figure}[!t]
\includegraphics[width=\linewidth]{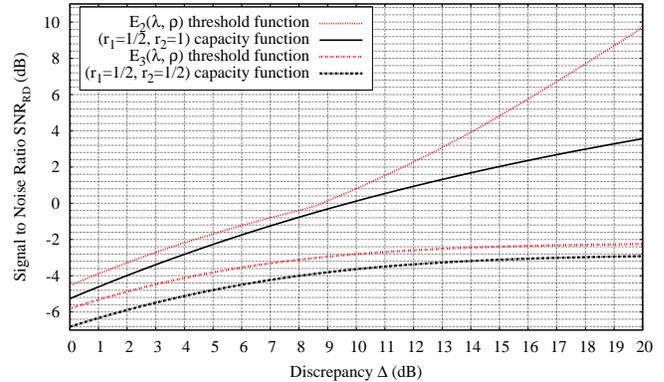}

\vspace{-3mm} \caption{Threshold vs. capacity function}\label{fig:se_ldpc_ex}\vspace{-5mm}
\end{figure}

First, we consider the SE-LDPC code ensemble ${\cal E}_2(\lambda, \rho)$. The designed distributed rate is $(r_1,r_2) =
(1/2, 1)$, meaning that the source broadcasts a code with rate $1/2$, and the relay generates a number of extended-bits
equal to the number of information bits. If the standard deviation of the white Gaussian noise on the source-to-relay
link is less than the above threshold $\sigma^{*}$ or, equivalently, the signal to noise ratio is greater than
$\mbox{SNR}^{*} = -2.70$ dB\footnote{Note that the theoretical limit is at $-2.82$ dB.}, we can assume error free
transmission between source and relay. From our definition of the discrepancy, the signal to noise ratios on the
source-to-destination and relay-to-destination links, are related by 
  $\mbox{SNR}_\RD = \mbox{SNR}_\SD + \Delta,$
where $\Delta = 10\log_{10}(\delta^2)$ is the discrepancy value in dB. The ensemble threshold and the capacity
functions are plotted at Figure \ref{fig:se_ldpc_ex}. We can observe that the gap between the two curves is relatively
small (between $0.7$ and $0.3$ dB) for discrepancy values $\Delta\in[0, 8.5]$, and it begins to increase starting from
this point. Hence, if a discrepancy value $\Delta > 8.5$~dB is not likely to be encountered in practice, the above
SE-LDPC code can be used to achieve reliable communication for channel parameters $(\delta,\sigma)$ very close to the
capacity.

\noindent {\em [Remark]} HARQ systems with incremental redundancy represent another possible application of the
proposed SE-LDPC codes. In this case extended-bits are transmitted by the source as incremental redundancy, whenever
the destination fails to decode the originally received signal. In such a case, the discrepancy is expected to take on
relatively small values.

Figure \ref{fig:se_ldpc_ex}  shows also the threshold function for the SE-LDPC code ensemble ${\cal E}_3(\lambda,
\rho)$, whose designed distributed rate is $(r_1,r_2) = (1/2, 1/2)$. We can observe that the gap between the threshold
and the capacity curves is between $1$ and $0.4$ dB for discrepancy values $\Delta\in[0, 20]$. This proves that
split-extending good codes for point-to-point communications results in good distributed codes for cooperative
communications.

\section{Conclusions} \label{sec:conclusions}
We proposed a new code-design method for LDPC coded cooperation, which is based on a split-and-extend approach. First,
we showed that the proposed design can be advantageously applied to existing codes, which allows for addressing
cooperation issues for evolving standards. Subsequently, we introduced the concepts of threshold and capacity
functions, and we derived density evolution equations for split-extended codes. Some ensemble thresholds have been
presented, showing that codes optimized for point-to-point communications can be split-extended, so that the
corresponding distributed codes perform close to the capacity of the relay channel for a wide range of discrepancy
values. Optimization of split-extended codes will be addressed in future works. Finally, besides advantageous
applications for cooperative transmission systems, the proposed design can also be used for communication systems
employing HARQ schemes with incremental redundancy.

 \bibliographystyle{../bib/IEEEbib}
\bibliography{../bib/MyBiblio,../bib/Zotero}

\end{document}